# Effect of Grain Boundaries on Thermal Transport in Graphene

Andrey Y. Serov,[1,2] Zhun-Yong Ong[2,3], and Eric Pop[1,2,4*]

[1]*Dept. of Electrical & Computer Eng., Univ. Illinois, Urbana-Champaign, IL 61801, USA*

[2]*Micro and Nanotechnology Laboratory, Univ. Illinois, Urbana-Champaign, IL 61801, USA*

[3]*Dept. of Physics, Univ. Illinois, Urbana-Champaign, IL 61801, USA*

[4]*Beckman Institute, University of Illinois, Urbana-Champaign, IL 61801, USA*

We investigate the influence of grain boundaries (GBs), line defects (LDs), and chirality on thermal transport in graphene using non-equilibrium Green's functions. At room temperature the ballistic thermal conductance is ~4.2 GWm$^{-2}$K$^{-1}$, and single GBs or LDs yield transmission from 50-80% of this value. LDs with carbon atom octagon defects have lower thermal transmission than GBs with pentagon and heptagon defects. We apply our findings to study the thermal conductivity of polycrystalline graphene for practical applications, and find that the type and size of GBs play an important role when grain sizes are smaller than a few hundred nanometers.

[*]Contact: epop@illinois.edu

Graphene is a promising material for applications as transparent electronics on flexible substrates,[1] or for high current density interconnects on common Si substrates.[2] Both in such practical contexts and from a fundamental point of view it is important to understand its thermal properties in order to address any arising thermal challenges.[3] For instance, exfoliated and suspended (monocrystalline) graphene has very high in-plane thermal conductivity,[4,5] comparable to or higher than that of diamond (2000-4000 $Wm^{-1}K^{-1}$). However, graphene grown by chemical vapor deposition (CVD) is more likely to be used for large-scale applications or interconnects,[1,2] but it is polycrystalline with grain boundaries (GBs)[6] and line defects (LDs)[7] which, as any imperfections,[8] can lower thermal conductivity.[9] Previous studies have examined heat flow across graphene GBs using non-equilibrium molecular dynamics (MD) simulations.[10,11] These enable the atomistic study of heat flow and the extraction of thermal conductivity based on computed temperature gradients,[12] taking into account anharmonic effects. However, the MD technique itself is based on classical equations of motion and may not capture quantum aspects of thermal transport, overestimating thermal conductance when operating temperatures are well below the Debye temperature, which is the case for graphene ($\Theta_D \sim 2100$ K).[3,5]

In this study we calculate the thermal conductance across GBs and LDs in graphene and in graphene nanoribbons (GNRs) using the non-equilibrium Green's function method (NEGF)[13] and compare our results with other studies and experimental data. We also find that chirality can have a significant impact on heat flow along GNRs, with some chiralities exhibiting thermal conductance significantly lower than that of armchair and zigzag structures, even without taking into account edge roughness effects. Then we compute the thermal conductance in GNRs with GBs and LDs. Finally, using our calculated GB transmission, we estimate the thermal conductivity of polycrystalline graphene depending on grain size for practical applications.

Figure 1 shows the different GBs and LD types considered in this work, obtained by using mirror reflection of GNRs with different chiralities. After establishing the structure of a GNR bisected by one GB, we find the minimum energy configuration using the MD simulator LAMMPS[14] with optimized Tersoff potentials proposed by Lindsay and Broido, shown to improve the accuracy of thermal calculations.[15] We use periodic boundary conditions during this minimization such that the resulting structure is less affected by GNR edges and is closer to the original unperturbed graphene lattice. The first GB shown in Fig. 1(a) is obtained with the GNR elementary cell lattice vector (1,4) i.e. $r = 4 \cdot n + 1 \cdot m$, where $n$ and $m$ are the basis vectors of the graphene lattice, as shown in Fig. 1(e); we call this particular GB "grain-4". We obtain additional GBs denoted "grain-7" and "grain-10" through a similar technique with elementary cell lattice vectors (1,7) and (1,10), respectively, as shown in Fig. 1(b-c). The LD shown in Fig. 1(d) has armchair lattice on both sides of the defect. The considered GBs consist of heptagons and pentagons along with ordinary carbon hexagons, while the LD consists of octagons and pentagons, all being consistent with recent experimental observations.[6,8,16]

The thermal conductance per unit cross-sectional area along a GNR can be written using a Landauer-like approach as[13]

$$G''(T) = \frac{1}{sW} \int\limits_{0}^{\infty} \frac{df(\omega)}{dT} \Xi(\omega) \hbar \omega \frac{d\omega}{2\pi}, \qquad (1)$$

where $W$ is the ribbon width, $s = 3.35$ Å is the graphene "thickness", $f(\omega)$ is the Bose-Einstein occupation factor, $\Xi(\omega)$ is the transmission function of phonons with frequency $\omega$, and $T$ is the lattice temperature. The transmission function $\Xi(\omega) = M_m T_{GB}$ is the product of the number of modes and the transmission coefficient of the GB or LD, if one is present.[17] (In the ballistic case without a GB or LD the transmission coefficient $T_{GB} = 1$.)

We calculate the transmission function using NEGF[13] with the force constant matrix based on the same optimized inter-atomic potentials[15] used for energy minimization when defining our structures. NEGF is a widely used method to calculate the phonon transmission properties of nanoscale carbon structures.[18] The force constant matrix is found as $k_{ij} = \partial^2 E / \partial u_i \partial u_j$, where $E$ is the potential energy of the lattice and $u_i$ is the displacement of the $i$-th degree of freedom.[13] In order to find the transmission function $\Xi(\omega)$ we need to calculate the Green's functions of semi-infinite GNRs to the left and right of the GB, which are found using force constant matrices at the GNR leads with the decimation technique.[19] We add the small imaginary part (0.1 percent) to the phonon frequency to achieve the convergence of the decimation technique. However, the sensitivity of the transmission function to the exact value of this imaginary part of frequency is weak. The equations used to calculate the Green's functions of the structure coupled to the electrodes and final transmission are standard and can be found elsewhere,[13,18] but for completeness are also reproduced in the supplement.[20] GNRs with GBs should be longer than 10 nm in length, otherwise the stress induced by the GB affects the lattice at the GNR end, thereby affecting the Green's functions of a pristine semi-infinite GNRs. The typical size of our simulated structures with GBs is approximately 11 nm in length and 6 nm in width. Although phonon-phonon scattering can be included in such a methodology,[21] we neglect it here in order to decrease computational burden, which can be justified because the intrinsic phonon mean free path at room temperature in pristine graphene[3] is ~600 nm in suspended samples and ~100 nm in substrate-supported samples, both being much larger than our structure size.

First we use periodic boundary conditions in the transverse direction ($y$-axis in Fig. 1) to define a force constant matrix of the system, which is related to the phonon propagation in the infinite graphene sheet. The corresponding phonon dispersion is shown in Fig. 2(a). We simulate a pristine GNRs with no defects ("nd" subscript) to ensure that the thermal conductance does not depend on chirality in this case of periodic boundary conditions as shown in Fig. 2(b). The transmission would be different if boundary conditions in the transverse direction were not periodic, which we will discuss later. We can see the significant contribution of out-of-plane flexural modes (Z-modes) at phonon energies below 50 meV, while in-plane modes (longitudinal L, and transverse T) are more important at higher energies. We simulated all structures using two different widths (~4 nm and ~6.5 nm) to ensure that the phonon transmission scales linearly with width, then used the wider samples (~6.5 nm) in our analysis.

Figure 2(c) shows the phonon transmission function of structures with the three GBs and LD from Fig. 1. We find that grain-4, grain-7 and grain-10 GBs have almost identical phonon transmission, and the LD exhibits the lowest transmission almost across the entire phonon spectrum. As the transmission is determined by the atomistic structure of the defect, the numerical results indicate that the prevalence of C-atom octagons have a greater effect on weakening thermal transport across the LD, than do the pentagons and heptagons in the other three GBs (see Fig. 1). In order to confirm this structural effect we evaluated the average cohesion energy ($E_{coh}$) of carbon atoms at the GB and LD boundaries and found that for LD it is -7.4 eV while for GBs it is in the range -7.63 to -7.77 eV, with the equilibrium value in pristine graphene being $E_{coh} =$ -7.97 eV (also see Table IV in Ref. 15). Higher deviation from the equilibrium energy implies a greater change in the force tensor, which leads to a greater change in transmission properties.

Having calculated the transmission functions, we can now obtain the thermal conductance by integrating equation (1) as shown in Fig. 3. The conductances of GBs and LD have different temperature dependencies because of their different transmission spectra. While at ~100 K the thermal conductances of all defects are very close (within 10 percent), at ~300 K the difference is notable. The conductance of GBs are similar[22] and reach ~80% of the pristine graphene conductance at 300 K; on the other hand the thermal conductance across the LD is lower, being ~50% that of pristine graphene at room temperature. Figure 3(b) displays each GB and LD conductance as a fraction of the non-defective (nd), pristine graphene conductance ($G''_{nd}$).

The conductances of defective structures in our calculations are in the range $G'' = 2\text{-}3$ GWm$^{-2}$K$^{-1}$ at room temperature. The conductance of pristine graphene is estimated to be around $G''_{nd} \approx 4.2$ GWm$^{-2}$K$^{-1}$. Thus, the thermal boundary conductance (TBC) owed to the defects alone is estimated as TBC = $1/\left(1/G'' - 1/G''_{nd}\right)$, which yields the range of 3-8 GWm$^{-2}$K$^{-1}$ at room temperature. These values are significantly lower than those obtained through MD simulations (> 15 GWm$^{-2}$K$^{-1}$).[10,11] While MD simulations are excellent tools to understand *relative* changes in thermal properties due to atomistic modifications,[11,23] they tend to overestimate the absolute value of the TBC due to their semi-classical treatment, in particular at temperatures well below the Debye temperature $\Theta_\mathrm{D}$, because of their inability to incorporate Bose-Einstein statistics.[3,5] Another difference between MD simulations and the NEGF approach is that MD simulations do take into account anharmonic phonon interactions, which are not captured in NEGF.

The Dulong-Petit classical limit[3,24] of our NEGF model is obtained at very high temperature ($T > \Theta_\mathrm{D}$), where thermal conductances saturate at ~11 GWm$^{-2}$K$^{-1}$ for pristine graphene, ~7 GWm$^{-2}$K$^{-1}$ for GBs, and ~5 GWm$^{-2}$K$^{-1}$ for the LD. The corresponding TBC values for GBs and LDs are 19 and 9 GWm$^{-2}$K$^{-1}$, respectively. The former, high-temperature NEGF estimate of the GB TBC, is in better agreement with the MD results of Bagri *et al.*[10] (15 to 45 GWm$^{-2}$K$^{-1}$) and those of Cao *et al.*[11] (~20 GWm$^{-2}$K$^{-1}$). However, the range of MD TBC results[10,11] remains higher than the high-temperature NEGF ones, suggesting that anharmonic interactions and multi-phonon processes (captured by MD) can enhance interfacial thermal transport.[25]

We now calculate the same structures without periodic boundary conditions in the transverse direction, which may be more representative of GBs across very narrow GNRs. Figure 4(a) shows that the ballistic thermal conductance along a zigzag GNR almost matches that of graphene with periodic boundary conditions, and the thermal conductance along an armchair GNR is lower, which was also demonstrated by other studies.[26] We find that GNRs with chiralities different from zigzag and armchair can have notably lower thermal transmission. We note that this reduction of phonon transmission shown in Fig. 4(a) is not due to the edge rough-ness scattering, but due to the reduction in number of modes ($M_m$) because phonon transmission in our simulations exhibited linear dependence on sample width.

It is interesting that the GNR with elementary lattice cell vector (1,10) has a structure very similar to a zigzag GNR (with angle between two orientations being only 4.6°), but almost 40 percent lower thermal conductance at 300 K because some phonon modes were lost as a result of omitting periodic boundary conditions in the transverse direction. This result is different from that obtained for thermal conductivity with relaxation time calculations[27] where an armchair GNR had the lowest thermal conductivity, and this difference may be due to the angular depend-ence of the scattering relaxation time.

We now turn to how GBs will affect thermal transport in GNRs without periodic boundary conditions in the transverse direction. The lattice structure is exactly the same as in the case of periodic boundary conditions and the difference lies in the definition of the force constant matrices. In order to perform a more meaningful analysis and to observe relative changes, we plot the ratio of thermal transmission in a GNR with a defect to the thermal transmission of a pristine GNR as it is shown in Fig. 4(b). We can see that the difference in normalized conduct-ance between GBs and LD shown in Fig. 4(c) is smaller here than it was in the case of periodic boundary conditions in the transverse direction shown in Fig. 3. Overall, the absence of periodic modes causes both lower thermal conductance of pristine GNRs and smaller transmission through defects, which indicates that periodic modes (which do not exist in realistic, narrow GNRs) are transmitted more effectively through GBs.

Before concluding, we wish to calculate the thermal conductivity in realistic polycrystalline graphene interconnects using the transmissions of GBs we just obtained with non-periodic boundary conditions in the transverse direction. The thermal conductivity determined by each polarization [longitudinal acoustic (LA), transverse acoustic (TA) or flexural acoustic (ZA)] is

$$\kappa = \frac{1}{2} \int_0^{\omega_{max}} \tau(\omega) g(\omega) \left( \frac{d\omega}{dq} \right)^2 \frac{df(\omega)}{dT} \hbar \omega \, d\omega, \tag{2}$$

where the ½ factor is due to the 2-dimensional wave vector $q$, $\tau(\omega)$ is the relaxation time, and $g(\omega)$ is the density of states numerically evaluated using the phonon dispersion given by the optimized Tersoff potentials.[15]

The phonon relaxation times can be estimated using Matthiessen's rule as $\tau(\omega)^{-1} = \tau_u^{-1} + \tau_s^{-1} + \tau_{GB}^{-1}$, where $\tau_u(\omega)$ is the phonon-phonon umklapp scattering time, $\tau_s(\omega)$ is the scattering with a substrate and $\tau_{GB}(\omega)$ is the relaxation time due to GB scattering. Although the relaxation time approximation (RTA) for umklapp scattering might not lead to the most accurate results,[28] it gives us a reasonable estimate for thermal conductivity at 300 K, the case we will discuss here. The treatment of umklapp scattering for the ZA modes can be complicated, but it was suggested that the ZA modes are strongly affected by the substrate[29] so we neglect their umklapp scattering and use only GB and substrate scattering for the ZA modes. We use tabulated substrate scattering times calculated for an SiO$_2$ substrate.[29]

The dependence of the thermal conductivity of supported monocrystalline graphene on temperature is shown in Fig. 5(a) and compared with experimental data.[29] In accordance with Boltzmann transport equation (BTE) simulations[29] the main contribution comes from TA and LA modes because the ZA modes are suppressed by substrate scattering. To describe GB scattering we use the relaxation time approach similar to the Mayadas model[30]

$$\tau_{GB} = \left(\frac{\partial \omega}{\partial q}\right)^{-1} \ell_G \frac{T_{GB}(\omega)}{1 - T_{GB}(\omega)} \tag{3}$$

where $\ell_G$ is the average grain size and $T_{GB}$ is the transmission coefficient of GBs shown in Fig. 4(c). In our NEGF model we can only separate the in-plane transmission from the out-of-plane transmission [see Fig. 2(b)] and cannot separate the LA and TA modes, therefore we will consider the transmission coefficient $T_{GB}$ to be the same for both LA and TA modes.

After performing the integration of equation (2) we obtain the thermal conductivity of supported polycrystalline graphene at room temperature,[31][31][31][31][31][31][29][30][30] which we plot as a function of grain size $\ell_G$ in Fig. 5(b). We find that LDs cause the strongest degradation of thermal conductivity, which is due to their lower phonon transmission as shown earlier. We also find that the thermal conductivity is not significantly degraded at room temperature if polycrystalline graphene grain sizes are several microns, or larger. However, the thermal conductivity is lower if the average grain size is smaller than several hundred nanometers, becoming comparable to the phonon mean free path in supported graphene (~100 nm) at room temperature.[3]

In conclusion, we calculated the thermal conductance of several GBs and LD in graphene using non-equilibrium Green's functions, and found that "not all grains are created equal" from a thermal transport point of view. We have identified ballistic transport limits, and shown that GNRs with chirality different from armchair and zigzag exhibit ~30% lower thermal conductance. Single GBs lying across such GNRs decrease the thermal conductance by another 30-40% compared to pristine GNRs of corresponding chirality. Importantly, all our calculations obey ballistic thermal transport limits. Finally, we estimated the dependence of thermal conductivity in substrate-supported CVD graphene on grain type and grain size, finding that GBs play an important role when grain sizes become comparable to several hundred nanometers. Such findings are important for future applications of CVD-grown graphene as devices or interconnects on flexible substrates which typically have very low thermal conductivity.

We acknowledge financial support from the Nanoelectronics Research Initiative (NRI), National Science Foundation (NSF) CAREER grant ECCS 09-54423, and valuable discussions with Zuanyi Li.

**FIGURES**

**Single-column positioning:**

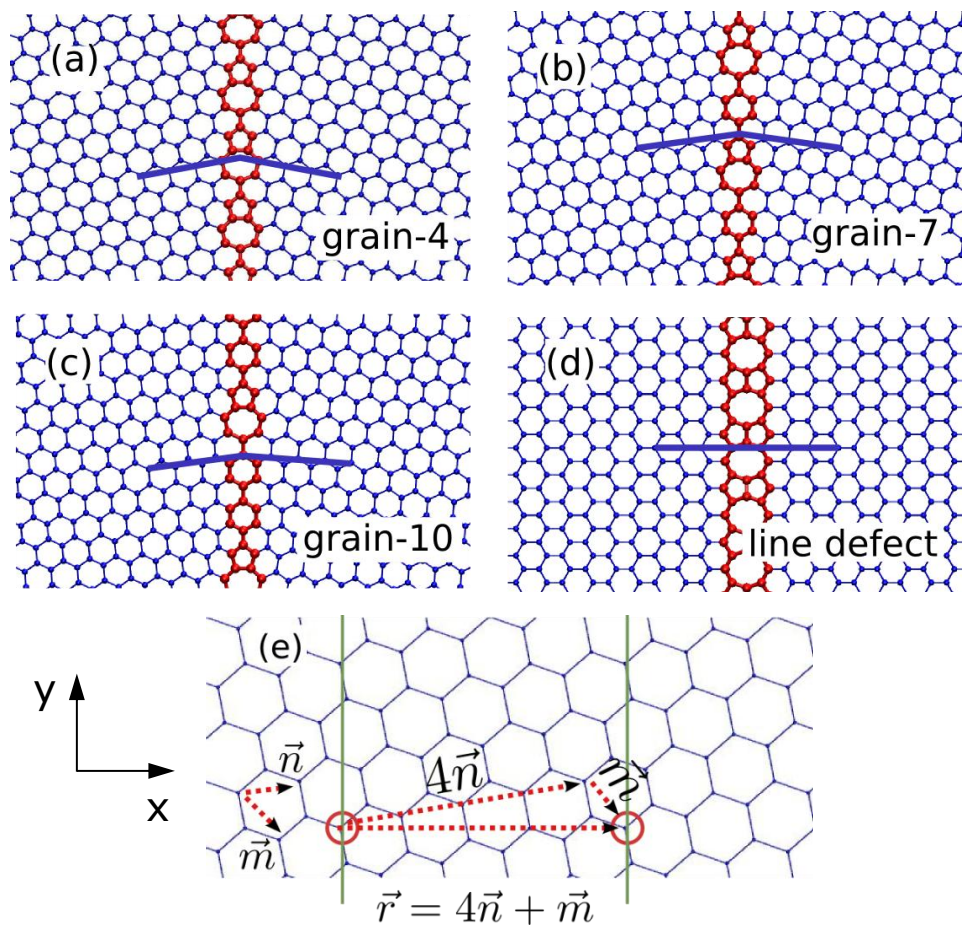

FIG. 1. (Color online) Examined graphene grain boundaries (GBs) constructed using mirror reflection of graphene nanoribbons (GNRs) with different chiralities: (a) grain-4, (b) grain-7, (c) grain-10 and (d) line defect (LD). (e) Example for the chirality of the GNR underlying grain-4.

**Single-column positioning:**

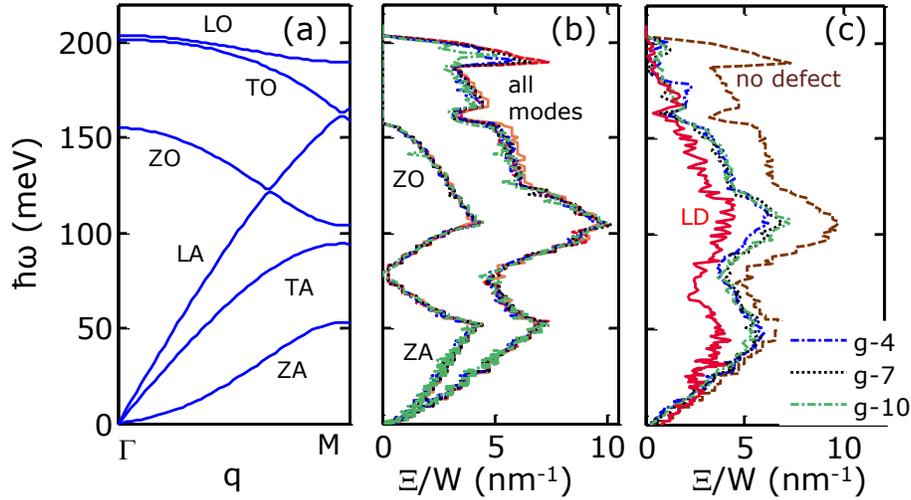

FIG. 2. (Color online) Phonon transmission along graphene structures with periodic boundary conditions in the transverse direction (also see Fig. 1). (a) Computed dispersion showing phonon energy $\hbar\omega$ vs. wave vector $q$. L, T and Z labels correspond to longitudinal, transverse, and out-of-plane phonon displacements. A and O labels are for acoustic and optical phonons, respectively. (b) Corresponding transmission function (i.e. number of modes per width) across pristine graphene of chiralities from Fig. 1. Individual chiralities are not labeled because all display the same transmission spectrum. The subset of out-of-plane ZA and ZO modes are shown separately. (c) Computed transport across grains from Fig. 1, revealing that transmission depends on the grain structure. Grain-4 (g-4) and grain-7 (g-7) have similar transmission, grain-10 (g-10) exhibits lower transmission, and line defect (LD) has the worst transmission (see text).

**Single-column positioning:**

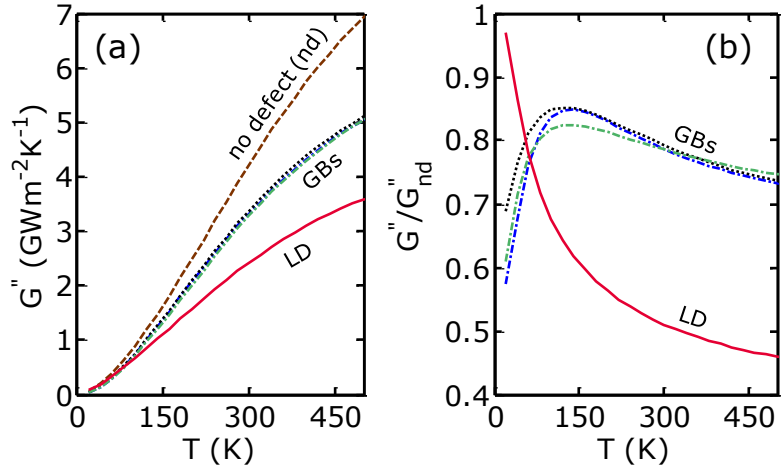

FIG. 3. (Color online) (a) Thermal conductance vs. temperature across various defects (GBs and LD) corresponding to Fig. 1. Calculations are performed using periodic boundary conditions in the transverse direction (width 6.5 nm) based on transmission spectra of Fig. 2. The upper limit of ballistic conductance in graphene with no defects ($G''_{nd}$) is displayed for comparison. (b) Thermal conductance vs. temperature along graphene with a defect, normalized by the ballistic conductance of the same case with no defects ($G''_{nd}$). At room temperature, the grain-4 and grain-7 GB structures show the largest thermal conductance (~80% of pristine graphene), and the LD the lowest (~50% of pristine graphene).

**Single-column positioning:**

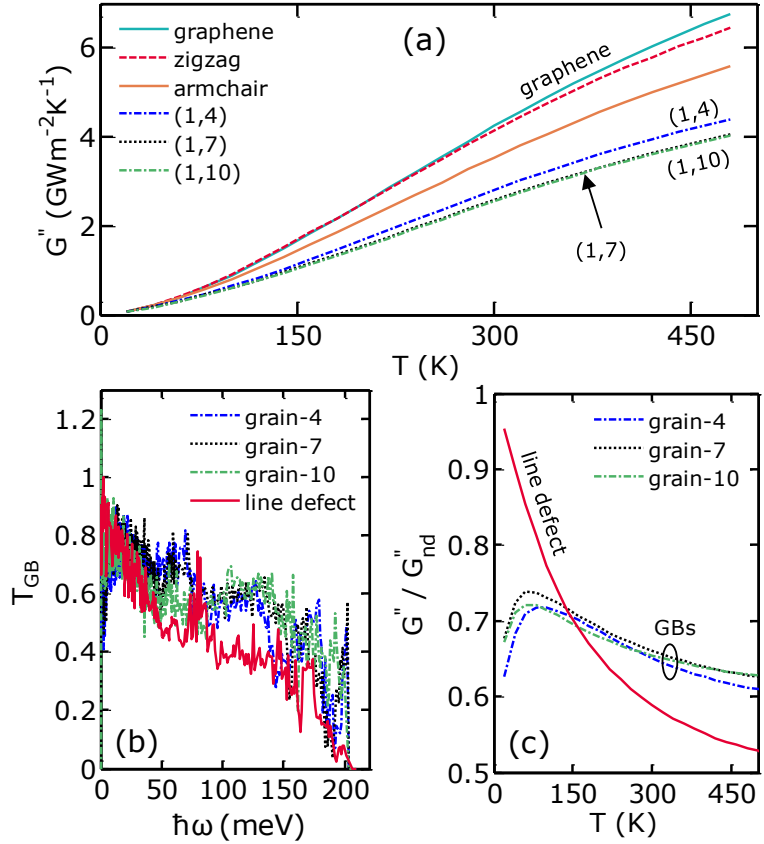

FIG. 4. (Color online) Thermal properties of structures calculated without using periodic boundary conditions in the transverse direction. (a) Ballistic thermal conductance in pristine GNRs of the chirality indicated, see Fig. 1. (b) Phonon transmission in GNRs with a GB or LD normalized by transmission of the same GNRs without defects ($T_{GB}$) as a function of phonon energy, $\hbar\omega$. (c) Thermal conductance vs. temperature along GNRs with a defect, normalized by the ballistic conductance of the same GNRs with no defects ($G''_{nd}$).

**Single-column positioning:**

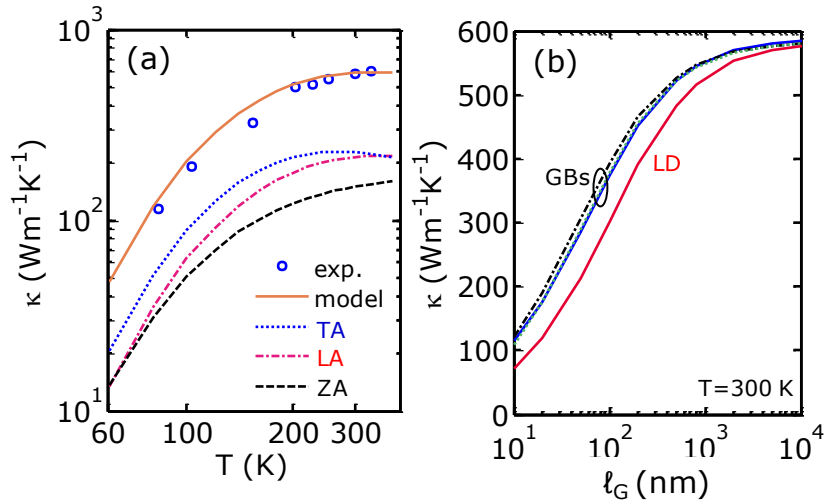

FIG. 5. (Color online) (a) Simulation results (without GBs, lines) fitted against experimental data (Ref. 29, symbols) for thermal conductivity of *mono*crystalline graphene on $SiO_2$ substrate. (b) Corresponding thermal conductivity of *poly*crystalline graphene as a function of average grain size $\ell_G$, calculated using the thermal conductance of GBs from Fig. 4. The thermal conductivity depends on defect type (GB or LD) and becomes strongly affected when grain sizes are below dimensions a few times the intrinsic phonon mean free path in substrate-supported graphene (~100 nm at room temperature) (also see Ref. 3).